\documentclass[twocolumn,english,pra]{revtex4}
\usepackage[T1]{fontenc}
\usepackage[latin9]{inputenc}
\usepackage{amsmath}
\usepackage{graphicx}
\usepackage{amssymb}
\usepackage{esint}

\makeatletter
\@ifundefined{textcolor}{}
{%
 \definecolor{BLACK}{gray}{0}
 \definecolor{WHITE}{gray}{1}
 \definecolor{RED}{rgb}{1,0,0}
 \definecolor{GREEN}{rgb}{0,1,0}
 \definecolor{BLUE}{rgb}{0,0,1}
 \definecolor{CYAN}{cmyk}{1,0,0,0}
 \definecolor{MAGENTA}{cmyk}{0,1,0,0}
 \definecolor{YELLOW}{cmyk}{0,0,1,0}
 }

\makeatother

\makeatother

\usepackage{babel}

\makeatother

\usepackage{babel}

\makeatother

\usepackage{babel}

\makeatother

\usepackage{babel}

\begin{document}

\title{Exact few-body results for strongly correlated quantum gases in two
dimensions}

\author{Xia-Ji Liu$^{1}$}

\email{xiajiliu@swin.edu.au}

\affiliation{$^{1}$ARC Centre of Excellence for Quantum-Atom Optics, Centre for
Atom Optics and Ultrafast Spectroscopy, Swinburne University of Technology,
Melbourne 3122, Australia}

\author{Hui Hu$^{1}$}

\email{hhu@swin.edu.au}

\affiliation{$^{1}$ARC Centre of Excellence for Quantum-Atom Optics, Centre for
Atom Optics and Ultrafast Spectroscopy, Swinburne University of Technology,
Melbourne 3122, Australia}

\author{Peter D. Drummond$^{1}$}

\email{pdrummond@swin.edu.au}

\affiliation{$^{1}$ARC Centre of Excellence for Quantum-Atom Optics, Centre for
Atom Optics and Ultrafast Spectroscopy, Swinburne University of Technology,
Melbourne 3122, Australia}

\date{\today{}}
\begin{abstract}
The study of strongly correlated quantum gases in two dimensions has
important ramifications for understanding many intriguing pheomena
in solid materials, such as high-$T_{c}$ superconductivity and the
fractional quantum Hall effect. However, theoretical methods are plagued
by the existence of significant quantum fluctuations. Here, we present
two- and three-body exact solutions for both fermions and bosons trapped
in a two-dimensional harmonic potential, with an arbitrary $s$-wave
scattering length. These few-particle solutions link in a natural
way to the high-temperature properties of many-particle systems via
a quantum virial expansion. As a concrete example, using the energy
spectrum of few fermions, we calculate the second and third virial
coefficients of a strongly interacting Fermi gas in two dimensions,
and consequently investigate its high-temperature thermodynamics.
Our thermodynamic results may be useful for ongoing experiments on
two-dimensional Fermi gases. These exact results also provide an unbiased
benchmark for quantum Monte Carlo simulations of two-dimensional Fermi
gases at high temperatures. 
\end{abstract}


\maketitle

\section{Introduction}

Two-dimensional (2D) strongly correlated quantum gases present unique
features from the point of view of many-body physics \cite{BlochRMP}.
Many sophisticated collective phenomena arise because of reduced dimensionality,
such as the long-sought Berezinsky-Kosterlitz-Thouless transition
\cite{theoryBKT1,theoryBKT2,exptBKT} and high-$T_{c}$ superconductivity
\cite{LeeRMP}. In addition, particles in 2D can have non-Abelian
quantum statistics, which is strikingly different from bosons and
fermions. For this reason, a 2D quantum system is a potential platform
for topological quantum computation in a way that is naturally immune
to decoherence \cite{NayakRMP}.

Recent experiments with ultracold atoms offer a unique opportunity
to investigate this physics in a controllable way \cite{BlochRMP,GiorginiRMP}.
In these experiments, one can modify aspects of the underlying geometry
and interactions between the atoms, at temperatures down to one billionth
of a degree above absolute zero. Experimental schemes to produce a
2D atomic quantum gas include a one-dimensional (1D) optical lattice,
formed by the superposition of two running laser waves \cite{Orzel,Burger,Kohl,MorschRMP,Spielman},
and strongly focused ellipsoidal optical traps. Using the technique
of Feshbach resonances \cite{ChinRMP}, the interatomic interaction
can also be changed from infinitely weak to infinitely strong. This
has already led to the observation of the crossover from a Bose-Einstein
condensate (BEC) to a Bardeen-Cooper-Schrieffer (BCS) superfluid in
three dimensions \cite{BlochRMP,GiorginiRMP}.

Theoretical investigations of 2D strongly correlated atomic quantum
gases, in particular the study of superfluidity in atomic Fermi gases,
have already attracted intense attention in the past few years \cite{BlochRMP,Randeria,Botelho,Petrov2003a,Martikainen,Bruun,Zhang,Tempere}.
However, theoretical methods for non-integrable 2D Fermi systems are
limited due to significant quantum fluctuations. Although a mean-field
approach combined with perturbation theory are usually adopted in
the understanding of the BCS-BEC crossover in three dimensions \cite{BlochRMP,GiorginiRMP,hldepl,hldnjp,natphys},
they may simply break down in 2D. Other traditional methods in condensed-matter
physics, such as exact diagonalization and quantum Monte Carlo simulation,
are often less helpful than one may expect, due to the restriction
to finite number of atoms or due to Fermi sign problems. Furthermore,
the harmonic trapping potential in ultracold atom experiments, which
is used to prevent the atoms from escaping, complicates theoretical
treatments.

In this paper, we present a few-particle perspective on strongly correlated
2D systems by exactly solving for the eigenstates of three identical
fermions or bosons in a 2D {\em isotropic} harmonic trap, with
arbitrary interaction strength. Three-fermion or three-boson problems
in three dimensions (3D) have been thoroughly investigated \cite{Braaten,WernerPRL,WernerPRA,Kestner},
covering many aspects such as the three-body recombination rate (or
stability) \cite{Petrov2003b,Petrov2004}, three-body perspective
on BEC \cite{Blume}, and Efimov physics \cite{Efimov,EfimovReview}.
The three-particle problem in low dimensions, however, is less well-studied
despite its considerable importance. There are very few studies of
universal low-energy properties of three identical bosons confined
in 2D \cite{Bruch,Nielsen,Hammer,Kartavtsev}.

Here, by constructing the exact wave functions, we solve and discuss
the full exact energy spectrum of three identical trapped fermions
or bosons in 2D. As the Efimov effect occurs only when the dimensionality
is greater than two \cite{Braaten}, all the states of fermions and
bosons that we study have {\em universal} properties determined
by a single parameter: the $s$-wave scattering length $a_{sc}$.
For three bosons, we find that an attractive interaction leads to
two distinct three-boson bound states in the form of a self-bound
boson droplet, as predicted by Hammer and Son \cite{Hammer} using
a 2D effective field theory.

Using few-particle exact solutions, we can also solve the problem
of a strongly correlated 2D quantum gas at high temperatures, including
both thermodynamics \cite{veprl} and dynamical properties \cite{vedsf,veaw},
using a quantum virial expansion method \cite{virialexpansion}. Here,
we calculate the second and third virial (expansion) coefficients
of a 2D Fermi gas. We then investigate the high-temperature equation
of state, including the chemical potential, energy and entropy, as
a function of temperature at a given interaction strength. Our thermodynamics
results give valuable insights for ongoing experiments on 2D Fermi
gases\cite{Dyke,Turlapov}. Further, these results may also provide
a useful benchmark for quantum Monte Carlo simulations for a 2D Fermi
gas at high temperatures, where convergence checks are otherwise difficult
to obtain.

The paper is organized as follows. In the next section, we present
exact solutions for the energy eigenstates of three-fermion and three-boson
systems with arbitrary $s$-wave interaction in an isotropic 2D harmonic
trap and discuss the resulting energy spectrum. In Sec. III, we calculate
the second and third virial coefficients of a 2D Fermi gas, at a given
temperature and interaction strength. Then, in Sec. IV, we investigate
the high-temperature thermodynamics of a strongly correlated 2D Fermi
gas. Sec. V is devoted to conclusions and final remarks. In the Appendix,
we outline some numerical details of the exact solutions.

\section{Exact few-particle solutions in a 2D harmonic trap}

We consider a 2D few-particle system of either fermions or bosons
in an isotropic 2D harmonic trap $V(\rho)=m\omega^{2}\rho^{2}/2$
with $\rho=\sqrt{x^{2}+y^{2}}$, where $\vec{\rho}_{j}=\left(x_{i},y_{j}\right)$
is the the $j$-th particle coordinate. For low-energy scattering,
the {\em attractive} interactions between atoms can be formally
described by a positive $s$-wave scattering length $a_{sc}$. For
identical fermions, there is no $s$-wave partial wave interaction
due to the Pauli exclusion principle. We thus consider for fermions
two different hyperfine (i.e., pseudo-spin) states, with the interaction
occurring only for two fermions with {\em unlike} spins. In the
case of a Feshbach resonance, which allows an adjustable interaction
strength, we focus on the case of a broad rather than narrow resonance.
This allows us to analyse the problem without considering an explicit
molecular channel. More generally, the molecular field causing the
resonance should be included, leading to a modified two-particle bound
state eigenfunction\cite{KheruntsyanDrummond2001}.

A peculiar feature of 2D interactions is that any attraction, whatever
how small, will support a two-particle bound state with binding energy
$E_{B}=4\hbar^{2}/[\exp\left(2\gamma\right)ma_{sc}^{2}]$, where $\gamma\simeq0.577216$
is the Euler constant \cite{Petrov2003a}. The interactions can then
be alternatively characterized by the two-particle binding energy
$E_{B}$. Contrary to the 3D BEC-BCS crossover situation, where the
bound state appears only at a certain interaction strength (i.e.,
unitarity limit), the scattering length $a_{sc}$ in 2D is always
{\em positive} due to the existence of a 2D bound state.

Following the idea introduced into two-body physics by Bethe and Peierls
\cite{BPBC}, it is convenient to replace the $s$-wave interaction
by a set of boundary conditions, which in 2D take the form \cite{Randeria,Adhikari,Petrov2001,Kartavtsev},
\begin{equation}
\lim_{\rho_{ij}\rightarrow0}\left[\rho_{ij}\frac{d}{d\rho_{ij}}-\frac{1}{\ln\left(\rho_{ij}/a_{sc}\right)}\right]\psi\left(\vec{\rho}_{1},\cdots,\vec{\rho}_{N}\right)=0,\end{equation}
 when particles $i$ and $j$ are close to each other. Here, $\psi\left(\vec{\rho}_{1},\cdots,\vec{\rho}_{N}\right)$
is the wave function of a system of $N$ particles and $\rho_{ij}=\left|\vec{\rho}_{i}-\vec{\rho}_{j}\right|$.
In addition to these Bethe-Peierls boundary conditions, the wave function
$\psi\left(\vec{\rho}_{1},\cdots,\vec{\rho}_{N}\right)$ satisfies
a non-interacting Schrödinger equation, \begin{equation}
\sum_{i=1}^{N}\left[-\frac{\hbar^{2}}{2m}{\bf \nabla}_{\rho_{i}}^{2}+\frac{1}{2}m\omega^{2}\rho_{i}^{2}\right]\psi=E\psi,\end{equation}
 with no two particles at the same coordinate.

\subsection{Two particles in a 2D harmonic trap}

As a preliminary study, let us first revisit the two-particle problem
\cite{Busch}. In a harmonic trap, the motion of the center-of-mass
${\bf C}=\left(\vec{\rho}_{1}+\vec{\rho}_{2}\right)/2$ can be separated
from the relative motion, and the relative Hamiltonian is given by,

\begin{equation}
{\cal H}_{rel}=-\frac{\hbar^{2}}{2\mu}{\bf \nabla}_{\rho}^{2}+\frac{1}{2}\mu\omega^{2}\rho^{2},\label{hamiRel2b}\end{equation}
 where $\vec{\rho}=\vec{\rho}_{1}-\vec{\rho}_{2}$ is the relative
coordinate and $\mu=m/2$ is the reduced mass. The energy level and
the corresponding wave function of two-particle system can be written
as $E=E_{cm}+E_{rel}$ and $\Psi_{2p}\left({\bf C},\vec{\rho}\right)=\phi_{2p}^{cm}\left({\bf C}\right)\psi_{2p}^{rel}\left(\vec{\rho}\right)$,
respectively. Here, the subscript {}``2p'' denotes the two-particle
problem.

The wave function of center-of-mass motion, $\phi_{2p}^{cm}\left({\bf C}\right)$,
is simply the well-known wave function of 2D harmonic oscillators
with $E_{cm}=(2n_{cm}+\left|m_{cm}\right|+1)\hbar\omega$, where the
good quantum number $n_{cm}$ and $m_{cm}$ label, respectively, the
number of nodes in the radial wave function and the angular momentum
of the center-of-mass motion. The relative wave function should be
solved by $\hat{H}_{rel}\psi_{2p}^{rel}\left(\vec{\rho}\right)=E_{rel}\psi_{2p}^{rel}\left(\vec{\rho}\right)$,
in conjunction with the Bethe-Peierls boundary condition, $\lim_{\rho\rightarrow0}[d/d\rho-1/(\rho\ln(\rho/a_{sc}))]\psi_{2p}^{rel}\left(\vec{\rho}\right)=0$.
The relative Hamiltonian has rotational symmetry and thus has a good
quantum number of angular momentum $m_{rel}$. Due to $s$-wave coupling,
it is easy to see that only the $m_{rel}=0$ branch of the relative
wave functions is affected by the interactions, so we focus on this
case.

We start by considering the solutions to the free Hamiltonian, without
including boundary conditions. The free relative Hamiltonian admits
two types of solutions, either in terms of the confluent hypergeometric
function of the first kind, $\exp(-\rho^{2}/2d^{2})_{1}F_{1}\left(-\nu,1,\rho^{2}/d^{2}\right)$,
or in terms of the Kummer confluent hypergeometric function of the
second kind, $\exp(-\rho^{2}/2d^{2})\Gamma(-\nu)U$$\left(-\nu,1,\rho^{2}/d^{2}\right)$,
where $d=\sqrt{\hbar/(\mu\omega)}$ is the length scale of the trap,
$\nu$ is determined by $E_{rel}=(2\nu+1)\hbar\omega$, and $\Gamma$
is the gamma function. The first kind of Kummer function $_{1}F_{1}$
is regular in the entire space and gives the standard wave function
of a 2D harmonic oscillator. In contrast, the second Kummer function
$U$ is singular at the origin.

Now, let us include the Bethe-Peierls boundary condition. It is easy
to see that one must choose the second type of Kummer solution as
the relative function, i.e., \begin{equation}
\psi_{2p}^{rel}\left(\vec{\rho}\right)\propto\exp(-\frac{\rho^{2}}{2d^{2}})\Gamma(-\nu)U\left(-\nu,1,\frac{\rho^{2}}{d^{2}}\right).\end{equation}
 The parameter $\nu$ or the relative energy $E_{rel}=(2\nu+1)\hbar\omega$
is then uniquely determined by the boundary condition. Considering
the property $\partial_{x}U\left(-\nu,1,x\right)=\nu U\left(1-\nu,2,x\right)$
and the asymptotic behavior of the confluent hypergeometric function
at $x\rightarrow0$,

\begin{eqnarray}
U\left(1-\nu,2,x\right) & = & -\frac{1}{\nu\Gamma\left(-\nu\right)x}+O\left(x^{0}\right),\\
U\left(-\nu,1,x\right) & = & -\frac{2\gamma+\ln x+\psi\left(-\nu\right)}{\Gamma\left(-\nu\right)}+O\left(x^{1}\right),\end{eqnarray}
 we immediately obtain the energy equation, \begin{equation}
\gamma+\frac{1}{2}\psi\left(-\nu\right)={\rm \ln}\left(\frac{d}{a_{sc}}\right).\label{energyRel2p}\end{equation}
 Here, $\gamma\simeq0.577216$ is the Euler constant and $\psi(x)$
is the digamma function.

\begin{figure}
\begin{centering}
\includegraphics[clip,width=0.4\textwidth]{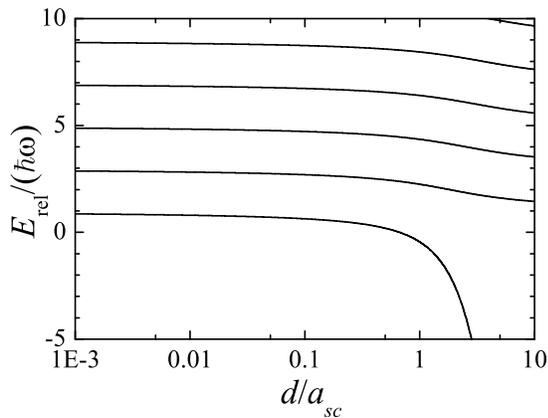} 
\par\end{centering}

\caption{(Color online). Relative energy spectrum of a two-particle system
with $m_{rel}=0$ as a function of the dimensionless interaction parameter
$d/a_{sc}$. The system goes to the strongly interacting limit when
$d/a_{sc}$ increases to an infinitely large value.}

\label{fig1} 
\end{figure}

In Fig. 1, we report the relative energy levels of a two-particle
system with $m_{rel}=0$ as a function of the dimensionless interaction
parameter $d/a_{sc}$. All the energy levels decrease with increasing
interaction strength, as expected for an attractively interacting
system. The lowest level corresponds to the ground state of a molecule
with size $a_{sc}$ and thus towards the strongly interacting limit
(i.e., $a_{sc}\rightarrow0$), it diverges as $-\hbar^{2}/(ma_{sc}^{2})$.
All the other excited levels instead converge to the non-interacting
limit.

It is interesting to note that with a positive scattering length,
the two particles interact repulsively if they do not occupy the ground
state of molecules. Thus, by excluding the lowest energy level, Fig.
1 can be alternatively viewed as the energy spectrum of two repulsively
interacting particles \cite{Duine}. Then, the right side with vanishing
$a_{sc}$ is the non-interacting limit for the repulsively interacting
system, and the unitarity limit of infinitely large $a_{sc}$ is the
strongly interacting limit.

In the limiting case of either zero or infinite scattering length,
one may calculate the asymptotic behavior of energy levels. We find
that, for the $n$-th level, the relative energy is given by, \begin{equation}
E_{rel}=\left[2n+1-\frac{2}{2\ln a_{sc}+\gamma+\sum_{k=1}^{n}1/k}\right]\hbar\omega,\end{equation}
 where $n=0,1,2,...$ is a non-negative integer and, in the limit
of $a_{sc}\rightarrow0$, the lowest molecule state has been excluded
in the count of energy levels, so that $n=0$ corresponds to the first
excited state.

\subsection{Three fermions in 2D harmonic trap}

Let us now turn to the three-particle problem. For three fermions,
we consider the configuration with two spin-up fermions (particle
1 and 3) and one spin-down fermion (particle 2), i.e., a $\uparrow\downarrow\uparrow$
configuration. It is convenient to use Jacobi coordinates. We define
the center-of-mass coordinate $\vec{\rho}_{CM}=\left(\vec{\rho}_{1}+\vec{\rho}_{2}+\vec{\rho}_{3}\right)/3$,
together with two relative coordinates ${\bf r}=\vec{\rho}_{1}{\bf -}\vec{\rho}_{2}$
and $\vec{\rho}=(2/\sqrt{3})\left[\vec{\rho}_{3}-\left(\vec{\rho}_{1}+\vec{\rho}_{2}\right)/2\right]$.
The solution for the center-of-mass motion is again the standard wave
function of a 2D harmonic oscillator. For the relative motion, on
top of the Bethe-Peierls boundary conditions, the relative Hamiltonian
reads \cite{WernerPRA}, \begin{equation}
{\cal H}_{rel}=-\frac{\hbar^{2}}{2\mu}{\bf \left(\nabla_{r}^{2}+\nabla_{\rho}^{2}\right)}+\frac{1}{2}\mu\omega^{2}\left({\bf r}^{2}+\rho{\bf ^{2}}\right).\label{hamiRel3f}\end{equation}

To solve the three-fermion problem, we extend the approach of Efimov\cite{Efimov}
to the trapped case and consider the following relative wave function
\cite{veprl}, \begin{equation}
\psi_{3f}^{rel}=\left(1-{\cal P}_{13}\right)\chi\left({\bf r},\vec{\rho}\right),\label{wfRel3f}\end{equation}
 where

\begin{equation}
\chi\left({\bf r},\vec{\rho}\right)=\sum_{n}a_{n}^{f}\psi_{2p}^{rel}({\bf r};\nu_{m,n})R_{nm}\left(\rho\right)\frac{e^{im\varphi}}{\sqrt{2\pi}},\end{equation}
 $R_{nm}\left(\rho\right)$ is the standard radial wave function of
2D harmonic oscillators with energy $(2n+|m|+1)\hbar\omega$, and
the set of parameters $\nu_{m,n}$ is determined by, \begin{equation}
E_{rel}=\left[\left(2n+|m|+1\right)+\left(2\nu_{m,n}+1\right)\right]\hbar\omega,\label{energyRel3f}\end{equation}
 for a given relative energy $E_{rel}$ and the two good quantum numbers
$n$ and $m$.

The wave function (\ref{wfRel3f}) is easy to understand. It is simply
a summation of products of the wave function of the paired fermions
(1 and 3), $\psi_{2p}^{rel}({\bf r};\nu_{m,n})$, and of the wave
function of particle 3 relative to the pair, $R_{nm}\left(\rho\right)e^{im\varphi}/\sqrt{2\pi}$.
The product certainly satisfies the relative Hamiltonian (\ref{hamiRel3f})
and gives rise to the energy conservation equation for $\nu_{m,n}$,
Eq. (\ref{energyRel3f}). Owing to the rotational symmetry of the
relative Hamiltonian, the angular momentum is well-defined and conserved.
In the relative wave function, we also include an exchange operator
for particle 1 and 3, which ensures the symmetry of the wave function
and ensures that the wave function satisfies the Pauli exclusion principle.
The set of coefficients $a_{n}^{f}$ can be uniquely determined using
the Bethe-Peierls boundary conditions. We note that because of the
exchange operator, the two boundary conditions reduce to just one,
since the other is satisfied automatically.

We now examine the Bethe-Peierls boundary condition which should lead
to a secular equation for the energy levels ($E_{rel}$) and wave
functions ($a_{n}^{f}$). Let us consider the first term, $\lim_{r\rightarrow0}r(d/dr)(1-{\cal P}_{13})\chi\left({\bf r},\vec{\rho}\right)$.
Recall that ${\cal P}_{13}\chi\left({\bf r},\vec{\rho}\right)=\chi\left({\bf r}/2-\sqrt{3}\vec{\rho}/2,-{\bf \sqrt{3}r}/2-\vec{\rho}/2\right)$,
which is regular at origin. Therefore, we find, \begin{eqnarray}
\lim_{r\rightarrow0}r\frac{d\psi_{3f}^{rel}}{dr} & = & \sum_{n}a_{n}^{f}R_{nm}\left(\rho\right)\frac{e^{im\varphi}}{\sqrt{2\pi}}\left[r\frac{d\psi_{2p}^{rel}}{dr}\right]_{r\rightarrow0},\\
 & = & \left(-2\right)\sum_{n}a_{n}^{f}R_{nm}\left(\rho\right)\frac{e^{im\varphi}}{\sqrt{2\pi}}.\label{BP3f1}\end{eqnarray}
 On the other hand, in the limit of $r\rightarrow0$, \begin{equation}
\frac{\psi_{3f}^{rel}}{{\rm \ln}\left(r/a_{sc}\right)}=\frac{\chi\left({\bf r},\vec{\rho}\right)-\chi\left(-\sqrt{3}\vec{\rho}/2,-\vec{\rho}/2\right)}{{\rm \ln}\left(r/a_{sc}\right)},\label{BP3f2}\end{equation}
 where effectively $\chi\left({\bf r},\vec{\rho}\right)_{r\rightarrow0}=\sum_{n}(-2)[\gamma+\psi(-\nu_{m,n})+\ln(r/d)]a_{n}^{f}R_{nm}\left(\rho\right)e^{im\varphi}/\sqrt{2\pi}$.
By substituting Eqs. (\ref{BP3f1}) and (\ref{BP3f2}) into the Bethe-Peierls
boundary condition, it is easy to show that, \begin{equation}
\sum_{n}a_{n}^{f}\left[B_{n}R_{nm}\left(\rho\right)+R_{nm}\left(\frac{\rho}{2}\right)\psi_{2p}^{rel}\left(\frac{\sqrt{3}\rho}{2};\nu_{m,n}\right)\right]=0,\end{equation}
 where \begin{equation}
B_{n}=\left(-1\right)^{m}2\left[\gamma+\psi(-\nu_{m,n})-\ln\left(\frac{d}{a_{sc}}\right)\right].\end{equation}
 The above equation can be solved by projecting the left-hand side
of the equation onto the expansion basis $R_{n^{\prime}m}\left(\rho\right)$,
which is orthogonal and complete. This leads to the secular equation,
\begin{equation}
\sum_{n^{\prime}}A_{nn^{\prime}}^{f}a_{n}^{f}=\ln\left(\frac{d}{a_{sc}}\right)a_{n}^{f},\end{equation}
 where the matrix elements are \begin{equation}
A_{nn^{\prime}}^{f}\equiv\left[\gamma+\psi\left(-\nu_{m,n}\right)\right]\delta_{nn^{\prime}}+\frac{\left(-1\right)^{m}}{2}C_{nn^{\prime}},\label{amatRel3f}\end{equation}
 and \begin{equation}
C_{nn^{\prime}}\equiv\int\limits _{0}^{\infty}\rho d\rho R_{nm}\left(\rho\right)R_{n^{\prime}m}\left(\frac{\rho}{2}\right)\psi_{2p}^{rel}(\frac{\sqrt{3}\rho}{2};\nu_{m,n^{\prime}}).\end{equation}
 It is clear that $C_{nn^{\prime}}$ arises from the exchange operator
${\cal P}_{13}$. In the absence of $C_{nn^{\prime}}$, the secular
equation is identical in form to Eq. (\ref{energyRel2p}), except
for an additional degree of freedom which corresponds to the motion
of particle 3 relative to the paired fermions (particle 1 and 2).
It then describes an {\em un-correlated} three-fermion system of
a pair and a single particle.

To solve the secular equation, one must imposes a cut-off $n_{\max}$
for the number of expansion functions of $R_{nm}\left(\rho\right)$.
The accuracy of the numerical calculations can be improved by increasing
$n_{\max}$. The relative energy level $E_{rel}$ is then implicit
in the secular equation via $\nu_{m,n}$. In practice, for a given
relative energy level $E_{rel}$, we diagonalize the matrix ${\bf A}_{f}=\{A_{nn^{\prime}}^{f}\}$
to obtain all the possible interaction strengths $d/a_{sc}$ that
correspond to this relative energy. We then invert the relations $a_{sc}(E_{rel})$
to calculate the desired energy spectrum (levels) as a function of
the interacting strength $d/a_{sc}$. The main numerical effort is
to calculate the matrix elements $C_{nn^{\prime}}$. We outline the
details of this procedure in the Appendix.

We note that, in both the two and three body cases, there are non-interacting
solutions to the point-contact interaction Hamiltonian. There are
many functions that vanish when two particles are at the same point.
For the two-particle case, these are the $m>0$ states. For the three-particle
case, the situation is more complicated. An example as pointed out
by Werner and Castin\cite{WernerPRL}, is the Laughlin state:

\begin{gather*}
\psi=e^{-\sum_{i=1}^{3}r_{i}^{2}/d^{2}}\prod_{1\leq n<m\leq3}\left[\left(x_{n}+iy_{n}\right)-\left(x_{m}+iy_{m}\right)\right]^{|\eta|}\end{gather*}
These states are not included in our interacting solutions.

\subsection{Three bosons in 2D harmonic trap}

For three bosons we can construct a similar relative wave function
to Eq. (\ref{wfRel3f}). This takes the form, \begin{equation}
\psi_{3b}^{rel}=\left(1+{\cal P}_{13}+{\cal P}_{23}\right)\chi\left({\bf r},\vec{\rho}\right),\label{wfRel3b}\end{equation}
 where

\begin{equation}
\chi\left({\bf r},\vec{\rho}\right)=\sum_{n}a_{n}^{b}\psi_{2p}^{rel}({\bf r};\nu_{m,n})R_{nm}\left(\rho\right)\frac{e^{im\varphi}}{\sqrt{2\pi}}.\end{equation}
 Compared with the fermion case, the only difference in the relative
wave function is that we need to include two exchange operators with
positive sign to enforce the proper symmetry of the bosonic wave function
\cite{WernerPRL}. This modifies the Bethe-Peierls boundary condition
and hence the secular equation. Otherwise, we follow the same derivation
as in the fermion case. By using ${\cal P}_{23}\chi\left({\bf r},\vec{\rho}\right)=\chi\left({\bf r}/2+\sqrt{3}\vec{\rho}/2,\sqrt{3}{\bf r}/2-\vec{\rho}/2\right)$,
we find that the secular matrix ${\bf A}_{b}=\{A_{nn^{\prime}}^{b}\}$
takes the form, \begin{equation}
A_{nn^{\prime}}^{b}\equiv\left[\gamma+\psi\left(-\nu_{m,n}\right)\right]\delta_{nn^{\prime}}+\left(-1\right)^{m+1}C_{nn^{\prime}},\label{amatRel3b}\end{equation}
 which has the same structure as $A_{nn^{\prime}}^{f}$. The difference
is that due to the additional exchange operator and different sign
before operators. The prefactor in the $C_{nn^{\prime}}$ terms is
$\left(-1\right)^{m+1}$, instead of $\left(-1\right)^{m}/2$ as in
Eq. (\ref{amatRel3f}).

It is of importance that in two dimensions the three-particle bosonic
wave functions we have constructed are universal, in the sense that
all the three-boson properties are determined by the single two-body
scattering length \cite{Kartavtsev}. This is contrary to the case
in three dimensions where even in the zero-range-interaction limit,
the Thomas and Efimov effect \cite{Efimov}, results in a set of universal
three-boson bound states which are described by an {\em additional}
three-body regularization parameter \cite{Efimov}.

The absence of an Efimov phenomenon, however, does not imply the absence
of three-body bound states. In free space, exactly two three-boson
bound states appear in two dimensions with an arbitrary two-body $s$-wave
scattering length, in the form of boson droplets \cite{Hammer}. The
ground bound state has a binding energy $E_{B3}^{(0)}=16.522688(1)E_{B}$,
while one excited bound state has $E_{B3}^{(1)}=1.2704091(1)E_{B}$.
Here, $E_{B}$ is the two-particle binding energy discussed earlier.

\subsection{Energy spectrum}

We now discuss the resulting energy spectrum of three fermions or
three bosons. Typically, we set a cut-off $n_{\max}=128$ for the
number of radial wave functions $R_{nm}\left(\rho\right)$ kept in
the calculation. By doubling and halving the value of $n_{\max}$,
we have checked that the relative accuracy of energy levels is less
than $<10^{-6}$, except for the $m=0$ subspace for bosons, where
the appearance of two three-boson bound states significantly decreases
the numerical accuracy.

\begin{figure}
\begin{centering}
\includegraphics[clip,width=0.4\textwidth]{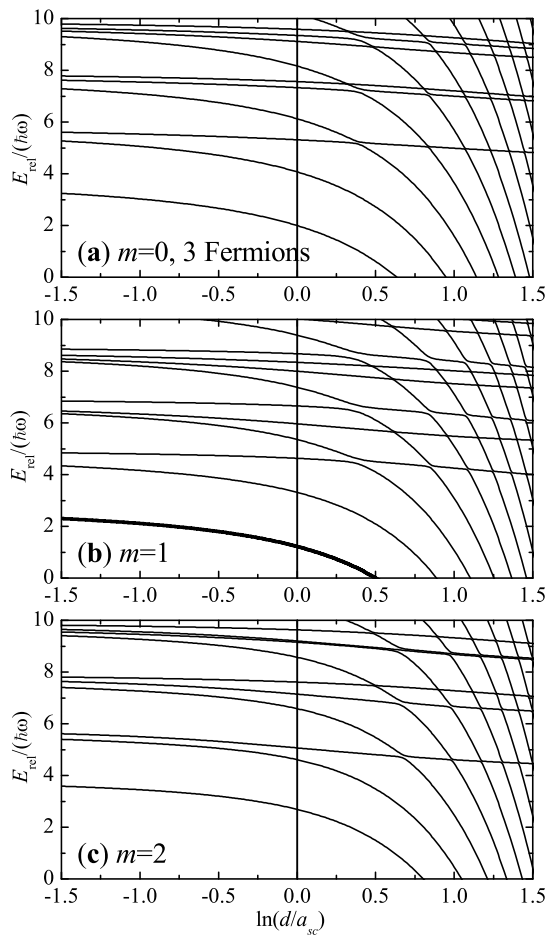} 
\par\end{centering}

\caption{(Color online). Relative energy spectrum of three trapped interacting
fermions in 2D, as a function of the dimensionless interaction parameter
$\ln(d/a_{sc})$. We show the spectrum in different subspaces of relative
angular momentum $m$. The ground state energy level in the subspace
$m=1$ has been highlighted by a thick line.}

\label{fig2} 
\end{figure}

\subsubsection{Three-fermion spectrum}

Fig. 2 gives the relative energy spectrum of a three-fermion system
at different relative angular momentum $m$ , as a function of the
interaction strength, $\ln(d/a_{sc})$. The ground state is in the
subspace $m=1$ due to the Pauli exclusion principle which prohibits
all three fermions from interacting when $m=0$, as highlighted by
a thick solid line. Compared with the two-body relative energy spectrum,
the energy levels are much more complicated. We observe two distinct
energy levels with decreasing scattering length and therefore increasingly
attractive interaction strengths. Some diverge to $-\infty$ as $a_{sc}^{-2}$,
while the others saturate to the limiting values that correspond to
the non-interacting energy spectrum. This essential feature exactly
resembles what we observed for the two-body relative energy spectrum
shown in Fig. 1, where the ground state of two particles diverges
to infinitely negative energy, while the other excited states converge
to the ideal, non--interacting spectrum. We note that the same feature
has also been observed very recently in calculations of a trapped
three-fermion system in 3D \cite{Daily}.

We may therefore identify the diverging energy level as the state
that contains a tightly bounded pair or molecule, together with a
fermion rotating around the molecule. The energy spacing of this {}``molecule
and atom'' state is roughly $2\hbar\omega$, accounting for the rotational
degree of freedom of the fermion. Accordingly, the other saturating
energy level is a state of three individual fermions, which therefore
should interact repulsively. In analogy to the two-particle case,
we may regard these {}``individual atom'' states as the energy states
of three {\em repulsively} interacting fermions with the same (positive)
$s$-wave scattering length, although there are necessarily many avoided-crossings
between the {}``molecule and atom'' states and the {}``individual
atom'' states. These appear particularly when the scattering length
$a_{sc}$ becomes comparable with the characteristic length scale
of the harmonic trap, $d$.

With this classification of energy levels in mind, the spectrum at
the limiting cases of $a_{sc}\rightarrow\infty$ and $a_{sc}\rightarrow0$
are easy to interpret. The former is simply the energy spectrum of
three weakly attractively interacting fermions, which, analogous to
the two-particle case, decrease linearly as $1/\ln(a_{sc})$ with
decreasing $a_{sc}$. The latter, excluding the {}``molecule and
atom'' states, is the spectrum of three weakly repulsively interacting
fermions, increasing linearly as $1/\ln(a_{sc})$ with increasing
$a_{sc}$. It is readily seen that in these two limiting cases the
energy levels, together with their degeneracy, are connected smoothly
with the spectrum of three ideal, non-interacting fermions.

\begin{figure}
\begin{centering}
\includegraphics[clip,width=0.4\textwidth]{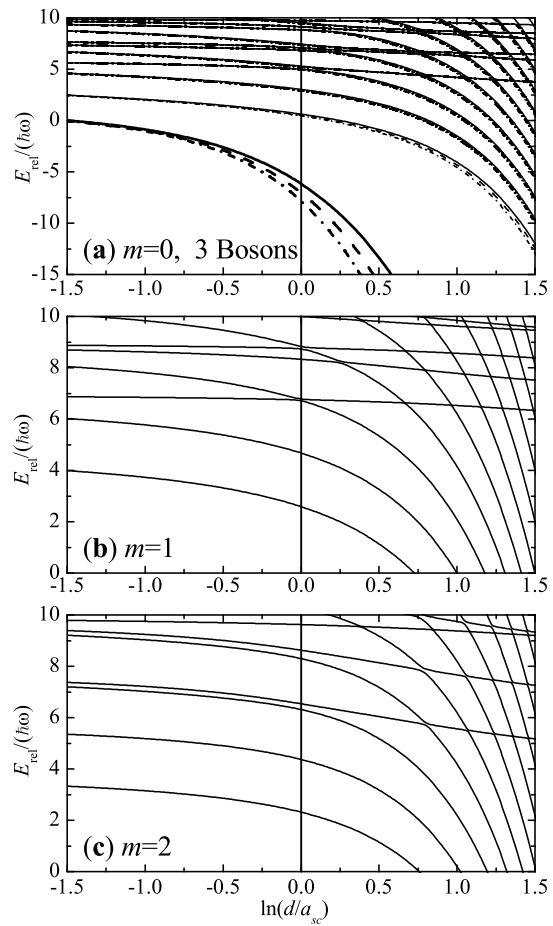} 
\par\end{centering}

\caption{(Color online). Relative energy spectrum of three trapped interacting
bosons in 2D at difference subspace, as a function of the dimensionless
interaction parameter $\ln(d/a_{sc})$. The ground state energy level
in the subspace $m=0$ has been highlighted by a thick line. The numerical
accuracy with $m=0$ is greatly suppressed due to the existence of
the self-bound droplet-like states. We thus plot the spectrum with
$n_{\max}=64$ (solid lines), $128$ (dashed lines), and $256$ (dot-dashed
lines), to show the slow convergence with respect to the number of
expansion basis elements $n_{\max}$.}

\label{fig3} 
\end{figure}

\subsubsection{Three-boson spectrum}

Fig. 3 presents the evolution of the relative energy spectrum of three
bosons with increasingly attractive interaction strength. In this
case, without the restriction of the Pauli exclusion principle, the
ground state is in the subspace of zero relative angular momentum,
$m=0$. We highlight this again by using a thick line. The essential
features of the spectrum are the same as in the spectrum for three
fermions. We observe both the {}``molecule and atom'' branch and
the horizontal {}``individual atom'' branch, together with some
avoided crossings between them. The latter branch may be viewed as
the spectrum of three repulsively interacting bosons.

However, there is an important difference, occurring in ground state
subspace with $m=0$. The lowest two states in the {}``molecule and
atom'' branch are three-boson bound states. One is the ground state
and the other is the lowest excited state. Their energy is lower than
the total energy of two attractively interacting bosons and a third
free-moving boson. In particular, the ground state energy is significantly
lower in magnitude than the two-body binding energy $E_{B}$. As a
result of these three-particle bound states, high numerical accuracy
is difficult to obtain. As shown in Fig. 3a, the energy levels of
the two bound states do not converge well even for the largest expansion
basis ($n_{\max}=256$) considered in these calculations.

The two bound states describe a self-bound bosonic droplet formed
via the attractive, short-ranged two-body potential, resembling the
well-known bright soliton of attractive bosons in 1D. Contrary to
the Efimov state, these bound states are universal and their properties
are determined entirely by the single $s$-wave scattering length.

We have estimated the binding energy of the two bound states at $\ln(d/a_{sc})=1$
by extrapolating the energy level obtained at a finite expansion basis
to $n_{\max}=\infty$. The interaction strength is chosen to minimize
the influence of the harmonic trap so that the size of the bound state
($\sim a_{sc}$) is much smaller the trapping scale ($\sim d$), while
at the same time to maintain the numerical result as accurate as possible.
Empirically, we find that the binding energy scales like, $E_{B3}(n_{\max})-E_{B3}\left(\infty\right)\propto n_{\max}^{-1/4}$.
This leads to $E_{B3}^{(0)}\simeq15.1E_{B}$ and $E_{B3}^{(1)}\simeq1.25E_{B}$,
which are reasonably in agreement with the accurate binding energies
in homogeneous space, $E_{B3}^{(0)}=16.522688(1)E_{B}$ and $E_{B3}^{(1)}=1.2704091(1)E_{B}$,
as predicted by a 2D bosonic effective field theory \cite{Hammer}.
The discrepancy, particularly for the ground state binding energy,
mainly comes from our insufficient numerical accuracy.

\section{Virial coefficients of strongly correlated fermions in 2D}

The knowledge of few-particle exact solutions provides a useful input
for investigating the high-temperature behavior of a strongly correlated
quantum gas, by applying a quantum virial expansion to the thermodynamic
properties \cite{veprl} or dynamical properties \cite{vedsf,veaw}.
Here, we are interested in the high-temperature equation of state
of strongly correlated fermions, which are now being accessed experimentally
in several laboratories.

The essential idea of the quantum virial expansion is that at high
temperatures where the chemical potential $\mu$ is strongly negative,
the fugacity $z\equiv\exp(\mu/k_{B}T)\equiv\exp(\beta\mu)\ll1$ is
a well-defined small parameter. We can therefore expand the thermodynamic
potential $\Omega$ of a quantum system in powers of the fugacity,
however strong the interaction strength is. Quite generally, we may
write \cite{veprl}, \begin{equation}
\Omega=-k_{B}TQ_{1}\left[z+b_{2}z^{2}+\cdots+b_{n}z^{n}+\cdots\right],\end{equation}
 where $b_{n}$ is the $n$-th (virial) expansion coefficient and
takes the following form, \begin{eqnarray}
b_{2} & = & \left(Q_{2}-Q_{1}^{2}/2\right)/Q_{1},\\
b_{3} & = & \left(Q_{3}-Q_{1}Q_{2}+Q_{1}^{3}/3\right)/Q_{1},\quad etc.\end{eqnarray}
 Here, $Q_{n}=Tr_{n}[\exp(-{\cal H}/k_{B}T)]$ is the partition function
of a cluster that contain $n$ particles and the trace $Tr_{n}$ is
taken over all the $n$-particle states of a proper symmetry. It is
clear that $Q_{n}$ and hence $b_{n}$ can be calculated once the
energy spectrum of up to $n$-body clusters is known. All the other
thermodynamic properties can then be derived from $\Omega$ via the
standard thermodynamic relations.

In a practical calculation, it is more convenient to focus on how
the virial coefficients are affected by interactions. We then may
consider the differences $\Delta Q_{n}=Q_{n}-Q_{n}^{(1)}$ and $\Delta b_{n}=b_{n}-b_{n}^{(1)}$,
where the superscript {}``$1$'' denotes an ideal, non-interacting
system having the same fugacity. As noted in the previous section,
our spectrum of the eigenstates does not include the non-interacting
solutions to the boundary value problem. We deal with this issue by
removing these states from both the interacting and non-interacting
summations that make up the trace differences $\Delta Q_{n}$. Since
they have the same energy with or without interactions, this does
not affect our results. Accordingly, we may rewrite the thermodynamic
potential in the form, \begin{equation}
\Omega=\Omega^{(1)}-k_{B}TQ_{1}\left[\Delta b_{2}z^{2}+\cdots+\Delta b_{n}z^{n}+\cdots\right],\label{omega}\end{equation}
 where $\Omega^{(1)}$ is the non-interacting thermodynamic potential
with the same fugacity and \begin{eqnarray}
\Delta b_{2} & = & \Delta Q_{2}/Q_{1},\\
\Delta b_{3} & = & \Delta Q_{3}/Q_{1}-\Delta Q_{2},\quad etc.\end{eqnarray}
 We now describe how to calculate the non-interacting thermodynamic
potential $\Omega^{(1)}$ and the virial coefficients $\Delta b_{n}$.

\subsection{Non-interacting thermodynamic potential $\Omega^{(1)}$}

Let us consider a two-component non-interacting Fermi gas in the thermodynamic
limit. In the limit of a large number of fermions, the non-interacting
thermodynamic potential $\Omega^{(1)}$is given {\em semiclassically}
by, \begin{eqnarray}
\Omega^{(1)} & = & -\frac{2}{\beta}\int\frac{d\vec{\rho}d{\bf k}}{\left(2\pi\right)^{2}}\ln\left[1+e^{-\beta\left(\frac{\hbar^{2}k^{2}}{2m}+\frac{m}{2}\omega^{2}\rho^{2}-\mu\right)}\right],\\
 & = & -2\frac{\left(k_{B}T\right)^{3}}{\left(\hbar\omega\right)^{2}}\int\limits _{0}^{\infty}t\ln\left(1+ze^{-t}\right)dt.\label{omega1}\end{eqnarray}
 Subsequently, the number of atoms, $N^{(1)}=-\partial\Omega^{(1)}/\partial\mu$,
and the entropy, $S^{(1)}=-\partial\Omega^{(1)}/\partial T$, may
be calculated, as well as the total energy, $E^{(1)}=\Omega^{(1)}+TS^{(1)}+\mu N^{(1)}$.
We find that, \begin{equation}
N^{(1)}=-2\left(\frac{k_{B}T}{\hbar\omega}\right)^{2}\int\limits _{0}^{\infty}t\frac{ze^{-t}}{1+ze^{-t}}dt\label{number1}\end{equation}
 and \begin{equation}
E^{(1)}=2\frac{\left(k_{B}T\right)^{3}}{\left(\hbar\omega\right)^{2}}\int\limits _{0}^{\infty}t^{2}\frac{ze^{-t}}{1+ze^{-t}}dt=-2\Omega^{(1)}.\end{equation}

\subsection{Second virial coefficient $\Delta b_{2}$}

We now calculate the second virial coefficient. We are interested
in the limit of a large number of fermions ($N\gg1$), a situation
that will mostly likely happen in experiment. As the Fermi energy
$E_{F}$ or the Fermi temperature $T_{F}=E_{F}/k_{B}$ is given by
$E_{F}=N^{1/2}\hbar\omega$ and the temperature $T\sim T_{F}$, we
shall define a reduced trapping frequency $\tilde{\omega}=\hbar\omega/k_{B}T\ll1$.
The thermodynamic limit is reached in the limit of $\tilde{\omega}\rightarrow0$
. In this limit, the single-particle partition function, determined
by the single-particle spectrum for a 2D harmonic oscillator $E_{nm}=(2n+\left|m\right|+1)\hbar\omega$
is given by $Q_{1}=2/(e^{+\tilde{\omega}/2}-e^{-\tilde{\omega}/2})\simeq2\left(k_{B}T\right)^{2}/\left(\hbar\omega\right)^{2}$,
which can also be determined from the first-order expansion of the
non-interacting thermodynamic potential $\Omega^{(1)}$. The prefactor
of two accounts for the two possible spin states of a single fermion.

The second virial coefficient $\Delta b_{2}$ is given by $\Delta Q_{2}$.
It is readily seen that the summation over the center-of-mass energy
in $Q_{2}$ gives exactly $Q_{1}/2$. Using the relative two-body
energy $E_{rel}=(2\nu_{n}+1)\hbar\omega$, where $\nu_{n}$ is $n$-th
solution of Eq. (\ref{energyRel2p}), we find that, \begin{equation}
\Delta b_{2}=\frac{1}{2}\sum_{\nu_{n}}\left[e^{-\left(2\nu_{n}+1\right)\tilde{\omega}}-e^{-\left(2\nu_{n}^{\left(1\right)}+1\right)\tilde{\omega}}\right],\label{db2}\end{equation}
 where the non-interacting $\nu_{n}^{\left(1\right)}=n$ ($n=0,1,2,...$)
is a non-negative integer.

\subsection{Third virial coefficient $\Delta b_{3}$}

The third virial coefficient, given by $\Delta b_{3}=\Delta Q_{3}/Q_{1}-\Delta Q_{2}$,
is more difficult to calculate. Both the term $\Delta Q_{3}/Q_{1}$
and $\Delta Q_{2}$ diverge as $\tilde{\omega}\rightarrow0$, with
the leading divergences canceling each other. We thus have to separate
out carefully the leading terms and treat them analytically. It is
easy to see that the spin configurations of $\uparrow\downarrow\uparrow$
and $\downarrow\uparrow\downarrow$ contribute equally to $Q_{3}$.
As $Q_{1}$ in the denominators cancels exactly with the summation
over the center-of-mass energy, we have $\Delta Q_{3}/Q_{1}=[\sum\exp(-E_{rel}/k_{B}T)-\sum\exp(-E_{rel}^{(1)}/k_{B}T)]$.
To calculate this, it turns out to be important to analyze the behavior
of $E_{rel}$ at large energies.

To this aim, we define a relative energy $\bar{E}_{rel}$, which is
the solution of Eq. (\ref{amatRel3f}) without the exchange term $C_{nm}$.
The utility of $\bar{E}_{rel}$ is that it can be constructed directly
from the two-body relative energy. In the subspace with a total relative
momentum $m$, it takes the form \begin{equation}
\bar{E}_{rel}=\left(2n+\left|m\right|+1\right)\hbar\omega+(2\nu+1)\hbar\omega,\label{ebar3f}\end{equation}
 where $\nu$ is the solution of the two-particle spectrum Eq. (\ref{energyRel2p}).
At large energies where the exchange effect becomes less important,
the full spectrum $E_{rel}$ approaches $\bar{E}_{rel}$ asymptotically.
There is an exception, however, at zero total relative momentum $m=0$.
The solution of $\bar{E}_{rel}$ at $n=0$ and $m=0$ is spurious,
due to the exchange operator which leads to a vanishing relative wave
function. It therefore cannot match any solution of $E_{rel}$. In
the $m=0$ subspace, we must require $n\geq1$ in Eq. (\ref{ebar3f}).

Interestingly, if we retain the spurious solution in the $m=0$ subspace,
the difference $[\sum\exp(-\bar{E}_{rel}/k_{B}T)-\sum\exp(-E_{rel}^{(1)}/k_{B}T)]$
gives $\Delta Q_{2}$ exactly, since the first part in Eq. (\ref{ebar3f})
is identical to the spectrum of center-of-mass motion. The spurious
solution gives the contribution, \begin{equation}
\sum_{\nu_{n}}\left[e^{-\left(2\nu_{n}+2\right)\tilde{\omega}}-e^{-\left(2\nu_{n}^{\left(1\right)}+2\right)\tilde{\omega}}\right]\equiv2e^{-\tilde{\omega}}\Delta b_{2},\end{equation}
 which should be subtracted. We thus finally arrive at the following
expression for the third virial coefficient, \begin{equation}
\Delta b_{3}=\sum\left[e^{-E_{rel}/k_{B}T}-e^{-\bar{E}_{rel}/k_{B}T}\right]-2e^{-\tilde{\omega}}\Delta b_{2}.\label{db3}\end{equation}
 The summation should be taken over all the possible relative energy
levels $E_{rel}$ and their asymptotic counterparts $\bar{E}_{rel}$.
It is well-behaved at arbitrary interaction strengths.

\subsection{Numerical results of virial coefficients}

We have numerically calculated the second and third virial coefficients
as functions of interaction strength and temperature, with a small
reduced trapping frequency $\tilde{\omega}\ll1$. To ensure the accuracy
of the calculations for $\Delta b_{3}$, we typically use a hundred
thousand relative energies $E_{rel}$. The dependence of the virial
coefficients on $\tilde{\omega}$ may be removed by a careful scaling
analysis.

\begin{figure}
\begin{centering}
\includegraphics[clip,width=0.4\textwidth]{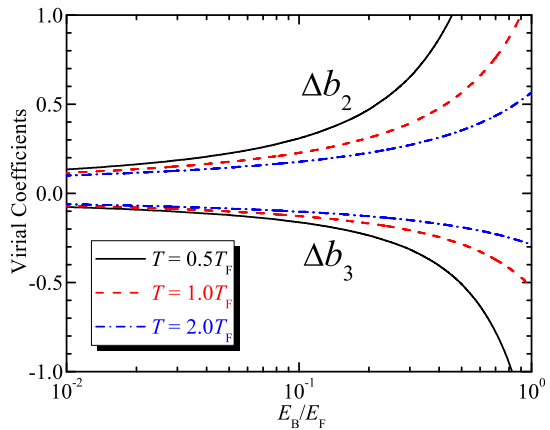} 
\par\end{centering}

\caption{(Color online). Second and third virial coefficients as a function
of the interaction strength $E_{B}/E_{F}$ at different temperatures,
$T/T_{F}=0.5$ (solid lines), $1.0$ (dashed lines), and $2.0$ (dot-dashed
lines).}

\label{fig4} 
\end{figure}

Fig. 4 shows the evolution of the virial coefficients with increasing
interaction strength, as characterized by the dimensionless two-body
binding energy $E_{B}/E_{F}$. The coefficients diverge exponentially
in the strongly attractively interacting limit, due to the formation
of tightly bound molecules. The lower the temperature, the faster
the divergence.

\begin{figure}
\begin{centering}
\includegraphics[clip,width=0.4\textwidth]{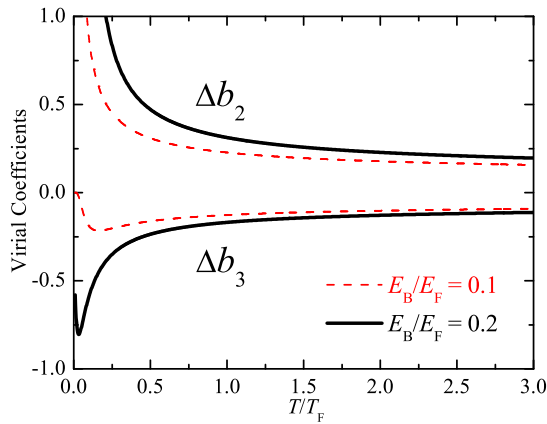} 
\par\end{centering}

\caption{(Color online). Temperature dependence of the second and third virial
coefficients at two interaction strengths, $E_{B}=0.2E_{F}$ (solid
lines) and $E_{B}=0.1E_{F}$ (dashed lines).}

\label{fig5} 
\end{figure}

Fig. 5 presents the temperature dependence of the virial coefficients
at two interaction strengths, $E_{B}=0.2E_{F}$ and $E_{B}=0.1E_{F}$.
The coefficients vary strongly with the temperature in the degenerate
regime ($T<T_{F}$). However, approaching the high-temperature Boltzmann
limit ($T\gg T_{F}$), the coefficients tend to saturate to a semiclassical
value.

\section{High-$T$ thermodynamics of strongly correlated fermions in 2D}

We are now in position to study the equation of state in the high
temperature regime. Using the thermodynamic relations, it is easy
to obtain,

\begin{equation}
N=N^{(1)}+2\left(\frac{k_{B}T}{\hbar\omega}\right)^{2}\left[2\Delta b_{2}z^{2}+3\Delta b_{3}z^{3}+\cdots\right],\label{number}\end{equation}
 and \begin{equation}
E=-2\Omega+2\frac{\left(k_{B}T\right)^{3}}{\left(\hbar\omega\right)^{2}}\frac{T}{T_{F}}\left[\Delta b_{2}^{\prime}z^{2}+\Delta b_{3}^{\prime}z^{3}+\cdots\right],\label{energy}\end{equation}
 where we have defined $\Delta b_{n}^{\prime}\equiv d(\Delta b_{n})/d(T/T_{F})$
and the Fermi temperature $T_{F}=\sqrt{N}\hbar\omega/k_{B}$. The
entropy is then calculated by using $S=(E-\Omega-\mu N)/T$, where
$\mu=k_{B}T\ln z$. Eqs. (\ref{omega}), (\ref{number}), and (\ref{energy}),
together with the non-interacting number equation (\ref{number1}),
form a closed set of expressions for thermodynamics.

We perform the calculation at a given fugacity within the trap units
$\hbar=m=\omega=k_{B}=1$. In the case of thermodynamic limit, the
temperature is fixed to an arbitrary constant (i.e., $T=100$). The
virial coefficients and their derivative with respect to the reduced
temperature are known as the input. We then calculate $N$ by using
the number equation (\ref{number}) with an initial guess of the reduced
temperature $T/T_{F}$ and obtain in turn the Fermi temperature $T_{F}=\sqrt{N}$.
The reduced temperature $T/T_{F}$ is updated. We iterate this procedure
until the final number of fermions and the reduced temperature converges
within a given relative error. We then calculate the total energy
using Eq. (\ref{energy}) and consequently the entropy $S=(E-\Omega)/T-Nk_{B}\ln z$.
We finally plot the chemical potential, entropy or energy per particle,
$\mu/E_{F}$, $S/(Nk_{B})$, and $E/(NE_{F})$, as a function of the
reduced temperature $T/T_{F}$.

\begin{figure}
\begin{centering}
\includegraphics[clip,width=0.4\textwidth]{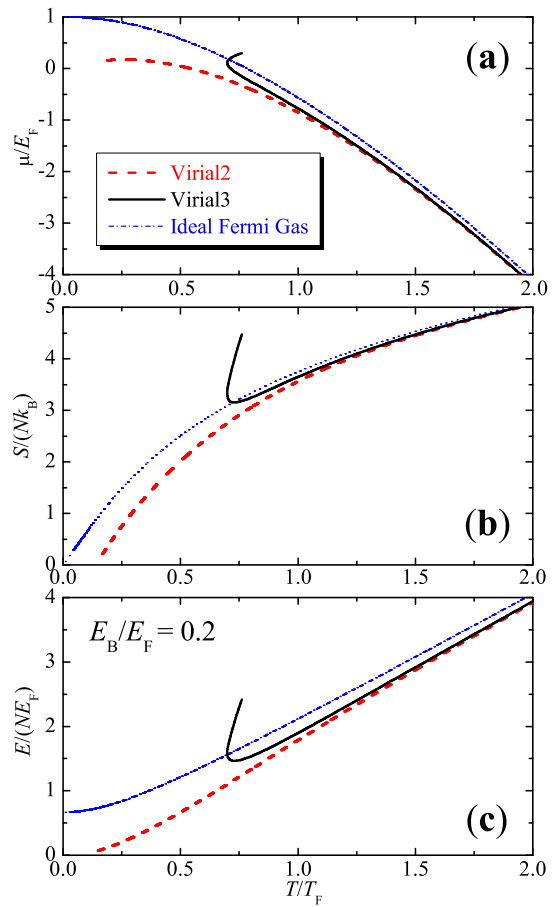} 
\par\end{centering}

\caption{(Color online). Temperature dependence of the chemical potential,
entropy, and energy of a strongly correlated Fermi gas in a 2D harmonic
trap. The predictions of virial expansion up to the third- and second-order
are shown, respectively, by the solid lines and dashed lines. For
comparison, we also show the ideal, non-interacting results using
dot-dashed lines.}

\label{fig6} 
\end{figure}

Fig. 6 gives the high-temperature equations of state of a strongly
correlated 2D Fermi gas at a typical interaction strength $E_{B}=0.2E_{F}$.
Compared with the ideal, non-interacting results, the equations of
state of a 2D trapped Fermi gas are strongly affected by interactions,
even in the high temperature regime. The applicability of the quantum
virial expansion method may be examined by comparing the prediction
of expansions of different orders. We estimate conservatively that
the third-order virial expansion is reliable down to the Fermi degeneracy
temperature, $T\sim T_{F}$.

\section{Conclusions and remarks}

In conclusion, we have presented the exact three-particle energy eigenstates
in a two-dimensional harmonic trap, for identical interacting fermions
and bosons. The energy spectra have been discussed in detail. We have
identified two types of energy levels, one containing a molecule and
the other consisting of individual atoms. The latter branch may be
interpreted as the energy spectrum of a repulsively interacting system.
For three strongly interacting bosons, we have found two universal
three-body bound states, corresponding to a self-bound boson droplet.
The calculated binding energy of the droplet is in reasonable agreement
with a previous theoretical prediction \cite{Hammer}.

Based on the these exact solutions, we are able to predict the high-temperature
thermodynamics of a strongly correlated quantum gas, by applying a
quantum virial expansion method. We have calculated for the first
time the second and third virial coefficients of a strongly correlated
two-dimensional Fermi gas in a harmonic trap and have calculated in
turn the temperature dependence of the chemical potential, entropy
and energy. Motivated by the striking experimental confirmation of
quantum virial expansion prediction for strongly interacting fermions
in three dimensions \cite{exptENS}, we anticipate that our prediction
in two dimensions will be tested in future experiments of two-dimensional
Fermi gases. Our thermodynamic results may also provide a useful benchmark
for future quantum Monte Carlo simulations at high temperatures for
two-dimensional systems of ultra-cold atoms. 
\begin{acknowledgments}
This work was supported in part by the ARC Centre of Excellence, ARC
Discovery Project Nos. DP0984522 and DP0984637, NSFC Grant No. 10774190,
and NFRPC Grant Nos. 2006CB921404 and 2006CB921306. 
\end{acknowledgments}
\appendix

\section{Calculation of $C_{nn^{\prime}}$}

In this Appendix, we outline the details of how to construct the matrix
element $C_{nn^{\prime}}$ in Eq. (\ref{amatRel3f}), which is given
by, \begin{equation}
C_{nn^{\prime}}\equiv\int\limits _{0}^{\infty}\rho d\rho R_{nm}\left(\rho\right)R_{n^{\prime}m}\left(\frac{\rho}{2}\right)\psi_{2p}^{rel}(\frac{\sqrt{3}}{2}\rho;\nu_{m,n^{\prime}}),\end{equation}
 where \begin{equation}
R_{nm}\left(\rho\right)=\sqrt{\frac{2n!}{\left(n+\left|m\right|\right)}}\rho^{\left|m\right|}e^{-\rho^{2}/2}L_{n}^{\left|m\right|}\left(\rho^{2}\right),\end{equation}
 is the radial wave function of an isotropic 2D harmonic oscillator
and the two-body relative wave function \begin{equation}
\psi_{2p}^{rel}=\Gamma(-\nu_{m,n^{\prime}})U(-\nu_{m,n^{\prime}},1,\frac{3}{4}\rho^{2})\exp(-\frac{3}{8}\rho^{2}).\end{equation}
 Here, for convenience we have set $d=1$ as the unit of length. $L_{n}^{\left|m\right|}$
is the generalized Laguerre polynomial and $U$ is the second Kummer
confluent hypergeometric function. A direct integration for $C_{nn^{\prime}}$
is difficult, since the second Kummer function becomes singular close
to the origin. Moreover, the integration for different values of $\nu_{m,n^{\prime}}$
makes the numerical calculation very time-consuming.

Thus, it is better to use a different strategy by writing, \begin{equation}
\psi_{2b}^{rel}=\sum_{k=0}^{\infty}\frac{1}{k-\nu_{m,n^{\prime}}}\frac{1}{\sqrt{2}}R_{k0}\left(\frac{\sqrt{3}}{2}\rho\right).\end{equation}
 Here, we have used the mathematical identity, \begin{equation}
\Gamma(-\nu)U(-\nu,1,x^{2})=\sum_{k=0}^{\infty}\frac{L_{k}\left(x^{2}\right)}{k-\nu}.\end{equation}
 Therefore, we arrive at \begin{equation}
C_{nn^{\prime}}=\sum_{k=0}^{\infty}\frac{1}{k-\nu_{m,n^{\prime}}}\frac{1}{\sqrt{2}}C_{nn^{\prime}k}^{m},\label{cnnp}\end{equation}
 where \begin{equation}
C_{nn^{\prime}k}^{m}\equiv\int\limits _{0}^{\infty}\rho d\rho R_{nm}\left(\rho\right)R_{n^{\prime}m}\left(\frac{\rho}{2}\right)R_{k0}\left(\frac{\sqrt{3}}{2}\rho\right)\end{equation}
 can be calculated with high accuracy by using an appropriate integration
algorithm. We note that, with a cut-off $n_{\max}$ for the number
of expansion functions (i.e., $n,n^{\prime}<n_{\max}$), $C_{nn^{\prime}k}^{m}$
vanishes identically for a sufficient large $k>k_{\max}\sim4n_{\max}$.
Thus, the summation over $k$ in Eq. (\ref{cnnp}) terminates naturally
and one does not need to worry about the convergence problem.

In the practical calculation, we tabulate and store the coefficients
$C_{nn^{\prime}k}^{m}$ in a file, for some given total relative angular
momentum $m$. Thus, the calculation of $C_{nn^{\prime}}$ for different
values of $\nu_{m,n^{\prime}}$ reduces to a simple summation, which
is very efficient and fast. We confirmed numerically that the matrix
$C_{nn^{\prime}}$ is symmetric, i.e., $C_{nn^{\prime}}=C_{n^{\prime}n}$.
A standard diagonalization algorithm can therefore be adopted for
the matrix ${\bf A}_{f}$ or ${\bf A}_{b}$.

\end{document}